\documentclass[twocolumn,aps,prb,superscriptaddress,showpacs]{revtex4}

\usepackage{amssymb}
\usepackage{amsmath}
\usepackage{amsfonts}
\usepackage{graphicx}

\begin{document}

\title{Suppressed spin dephasing for 2D and bulk electrons in GaAs wires\\
due to engineered cancellation of spin-orbit interaction terms}



\author{S.~Z.~Denega}
\author{T.~Last}
\author{J.~Liu}
\author{A.~Slachter}
\author{P.~J.~Rizo}
\author{P.~H.~M.~van~Loosdrecht}
\author{B.~J.~van~Wees}
\affiliation{Zernike Institute for Advanced Materials,
University of Groningen, NL-9747AG Groningen, The
Netherlands}
\author{D.~Reuter}
\author{A.~D.~Wieck}
\affiliation{Angewandte Festk\"{o}rperphysik, Ruhr-Universit\"{a}t
Bochum, D-44780 Bochum, Germany}
\author{C.~H.~van~der~Wal}
\affiliation{Zernike Institute for Advanced Materials,
University of Groningen, NL-9747AG Groningen, The
Netherlands}


\date{April 15, 2010}

\begin{abstract}
We report a study of suppressed spin dephasing for quasi-one-dimensional
electron ensembles in wires etched into a GaAs/AlGaAs
heterojunction system. Time-resolved Kerr-rotation
measurements show a suppression that is most pronounced for
wires along the [110] crystal direction. This is the
fingerprint of a suppression that is enhanced due to a
strong anisotropy in spin-orbit fields that can occur when
the Rashba and Dresselhaus contributions are engineered to
cancel each other. A surprising observation is that this
mechanisms for suppressing spin dephasing is not only
effective for electrons in the heterojunction quantum well,
but also for electrons in a deeper bulk layer.
\end{abstract}



\maketitle

\begin{figure}[b!]
\includegraphics[width=70mm]{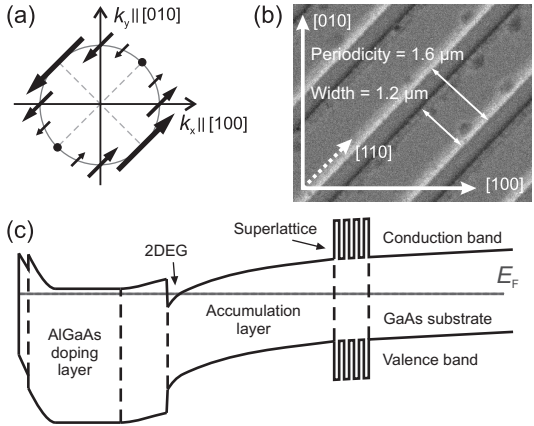}
\caption{(a)~Dependence of spin-orbit fields on $k$ vector on the
Fermi circle in the $k_x$-$k_y$ plane for equal strength of Rashba
and linear Dresselhaus effect. (b)~Scanning-electron micrograph for
a section of a wire array oriented along the [110] crystal
direction. (c)~Band diagram of the GaAs/AlGaAs heterostructure (not
to scale), with a 2DEG where the conduction band bends below the
Fermi energy $E_F$.}\label{fig:introfig}
\end{figure}

For electron ensembles in semiconductor device structures the
dominant mechanism for spin dephasing and spin relaxation is often
due to spin-orbit interaction (SOI). A current challenge in the
field of spintronics is therefore to obtain control over such
relaxation, or to even turn SOI into a resource for spin
manipulation.\cite{awschalom2009phys} GaAs/AlGaAs based
heterostructures provide here an interesting platform. Besides high
material quality and good optical selection rules, these materials
can be engineered to have two types of SOI: Dresselhaus SOI,
resulting from lack of inversion symmetry in the GaAs crystal
structure, and Rashba SOI, effective when electrons are confined in
a layer with bending of the conduction band.\cite{silsbee2004jpcm}
Such SOI acts as a $k$-vector-dependent effective magnetic field on
electron spins. Notably, the strength of these two effects can be
tuned to exactly cancel each other
\cite{miller2003prl,koralek2009nature} for electrons with the
$k$ vector in [110] direction [Fig.~\ref{fig:introfig}(a)]. This
notion is central in many proposals
\cite{averkiev1999prb,bernevig2006prl,schliemann2003prl,cheng2007prb,duckheim2007prb}
that aim at suppressing spin relaxation effects. Work on unpatterned
quantum-well areas with techniques that filter out signals from
electrons with [110] $k$ vectors confirmed reduced dephasing for
these electrons.\cite{averkiev2006prb,korn2008physE,koralek2009nature}

We report here the observation that this cancellation of two SOI
terms can directly result in suppressed spin dephasing for an entire
electron ensemble in a wire along the [110] direction. Notably, this
does not require that full quantum confinement restricts electron
transport in the wire to the [110] direction. Instead, it concerns a
strong enhancement of a motional-narrowing type effect that results
from frequent scattering on the boundaries of quasi-one-dimensional (1D) wires
\cite{holleitner2006prl,Kohda2009,frolov2009nature} that can---with
further tuning---suppress spin dephasing by several orders of
magnitude.\cite{liu2009jsnm,Kohda2010} Quasi-1D systems are here defined as
having the wire width too wide for 1D quantum confinement, but
smaller than the spin precession length (ballistic travel distance
during which spins precess an angle $\pi$ in the sum of spin-orbit
fields and external magnetic field). In addition, the wire width
should not be much in excess of the mean free path.

We used electron ensembles in GaAs/AlGaAs wires
(Fig.~\ref{fig:introfig}) that contain a heterojunction-based two-dimensional
electron gas (2DEG) for which the Rashba and linear Dresselhaus SOI
 are of similar strength.\cite{miller2003prl}
We used time-resolved Kerr readout for directly testing whether
quasi-1D confinement in device structures combined with anisotropy
for SOI in $k$ space can be tuned to give a pronounced suppression
for spin dephasing for wires in the [110] direction. For 2DEG
electrons we indeed observe this behavior, which we will further
call spin-dephasing anisotropy (SDA). Surprisingly, we also obtained
evidence for a new manifestation of SDA: a contribution to our Kerr
signals from a deeper bulk layer in our wires also shows SDA. We can
explain these observations by considering that an electron ensemble
in a bulk layer with weak band bending [Fig.~\ref{fig:introfig}(c)]
is also subject to a Rashba-type SOI and we find agreement with
numerical modeling of such a situation.


We fabricated arrays of quasi-1D wires with electron-beam
lithography and subsequent wet etching of a GaAs/AlGaAs
heterojunction system with a 2DEG at 114~nm depth (etch
depth $\sim$100~nm). Figure~\ref{fig:introfig}(b) is a
micrograph of an array with wires along [110] (1.2~$\mu$m
wire width and 1.6~$\mu$m periodicity). Other areas had [100],
[010], and \mbox{[-110]} oriented wires, and for reference
we had both unpatterned and fully etched areas.
Figure~\ref{fig:introfig}(c) shows a band diagram of the
GaAs/AlGaAs heterostructure along growth direction. The
layer sequence starts with an intrinsic (001) oriented GaAs
substrate, followed by a training layer (superlattice with
10 periods of 5.2~nm GaAs and 10.6~nm AlAs). Then a 933~nm
undoped GaAs accumulation layer is followed by a 37~nm
undoped Al$_{0.32}$Ga$_{0.68}$As spacer layer. On top of
this 72~nm homogeneously Si-doped Al$_{0.32}$Ga$_{0.68}$As
is grown. The heterostructure is finished by a 5~nm GaAs
cap layer. The 2DEG is at the heterojunction between the
AlGaAs spacer layer and the GaAs accumulation layer. Under
illuminated conditions at 4.2~K the electron density is
$n_{2D}=4.7\times10^{15} \; \rm{m}^{-2}$ and the mean free
path is $\sim$30~$\mu$m.

The spin dynamics was probed with conventional
\cite{kikkawa1998prl} mono-color time-resolved
Kerr-rotation measurements in an optical cryostat at 4.2~K
with magnetic fields (magnitude $B$) up to 7~T. We used a
tunable Ti:sapphire laser with $\sim$150~fs pulses at
80~MHz repetition rate. The spectrum of pulses was wider
than Fourier-transform limited, with significant amplitudes
over a $\sim$15~meV window. Samples were excited at normal
incidence with a circularly polarized pump pulse (helicity
modulated at 50~kHz with a photoelastic modulator for
avoiding nuclear polarization). This induces a
nonequilibrium electron ensemble in the conduction band
with a net spin orientation along the growth direction of
the sample. The evolution of the spin ensemble is recorded
by measuring the 50~kHz component in Kerr rotation after
reflection of a linearly polarized probe pulse with a
polarization bridge, as a function of pump-probe delay $t$.
We used spot diameters of about 100~$\rm{\mu m}$ on wire
arrays of larger dimensions.\cite{grating} All data were
taken at pump-photon density $n_{ex} \approx
3\times$10$^{15}$~m$^{-2}$ per pulse. Measurements were
performed with the external field in [110] direction,
unless stated otherwise.

\begin{figure}[t!]
\begin{center}
\includegraphics[width=80mm]{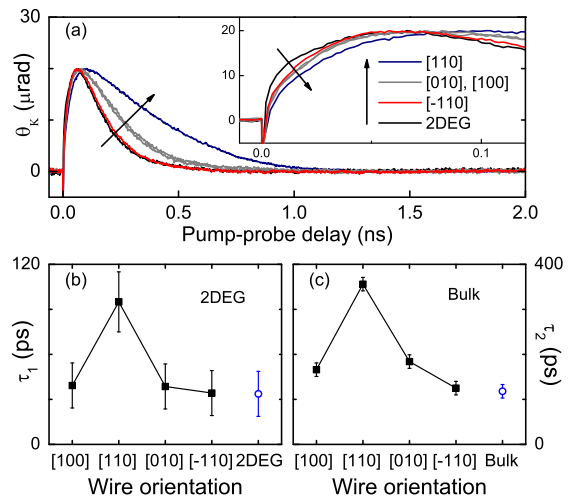}
\end{center}
\caption{(Color online) Spin-dephasing anisotropy at zero external magnetic field.
(a)~Kerr rotation as a function of pump-probe delay for the
unpatterned area and wires in \mbox{[-110]}, [110], [100], and [010]
directions (order along arrow in traces and legend match). The inset shows the Kerr response at short
delays to highlight the slow growth in all traces. [(b) and (c)]
Spin-dephasing times for both the 2DEG ($\tau_{1}$) and bulk ($\tau_{2}$)
spin populations.} \label{fig:sdazerofield}
\end{figure}

\begin{figure}[b!]
\begin{center}
\includegraphics[width=80mm]{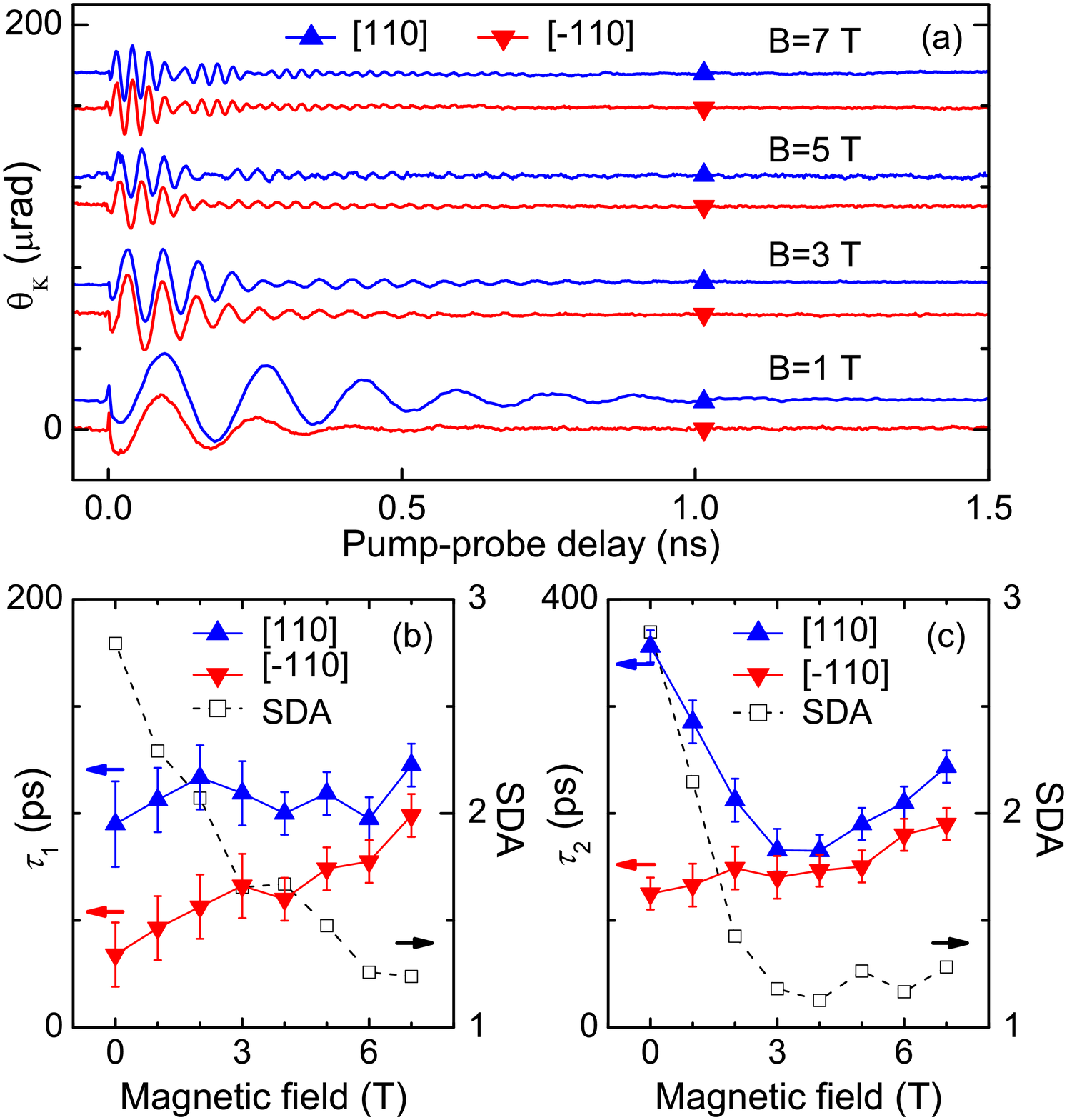}
\end{center}
\caption{(Color online) Evolution of spin-dephasing anisotropy (SDA) in transverse
magnetic fields up to $B=7$~T. (a) Kerr signals as a function of
pump-probe delay from [110] and \mbox{[-110]} oriented wires (traces
offset for clarity). [(b) and (c)] Corresponding spin dephasing times
$\tau_{i}$ as a function of $B$, for (b) 2DEG and (c) bulk
electrons. These plots also present SDA quantified as
$\tau_{i,[110]} / \tau_{i,[-110]}$.}\label{fig:sdainfield}
\end{figure}

While a heterojunction system is for our research an interesting
material given the strong and tunable band bending in the 2DEG
layer, it imposes difficulties for an experiment with optical
probing (as compared to double sided quantum wells). The bandgaps
for the 2DEG layer, and deeper parts of the accumulation layer and
the substrate are nearly identical [Fig.~\ref{fig:introfig}(c)], so
probing must use reflection and one cannot avoid that Kerr signals
have contributions from both 2DEG electrons and bulk electrons in
these deeper layers. We recently characterized \cite{rizo2009prb}
that it is for our material nevertheless possible to isolate the
Kerr signal from 2DEG electrons when pumping at least $\sim$20~meV
above the bottom of the conduction band for bulk (we used 1.55~eV).
In summary, at pump intensities with $n_{ex} \leq n_{2D}$ the Kerr
signal $\theta_K$ is a superposition of two mono-exponentially
decaying cosines,
\begin{equation}
\theta_K  = \sum_{i=1,2} A_i \exp \left(
-\frac{t}{\tau_{i}} \right) \cos \left( \frac{|g_i| \mu_B
B}{\hbar}t + \phi_i \right) \label{eq:kerrfitformula}
\end{equation}
with amplitudes $A_i$, $g$ factors $|g_i|$, dephasing times
$\tau_i$, and $\phi_i$ are apparent initial phases
\cite{phases} for spin precession. This also applies to the
data that we present here.
Figures~\ref{fig:sdazerofield}(a) and
\ref{fig:sdainfield}(a) show Kerr signals from wires,
recorded in fields of 0~T to 7~T. In particular at high
fields, the envelopes of oscillating Kerr signals show
clear nodes and anti-nodes and we checked that all Fourier
transforms of data taken at $B>3$~T showed two pronounced
peaks at $g$ factors $|g_1|\approx0.36$ and
$|g_2|\approx0.43$. Further, Eq.~\ref{eq:kerrfitformula}
always yields good fits for these same $g$-factors, also
when applied to data in the range 0~T to 3~T. In our
earlier work we identified the contribution with
$|g_1|\approx0.36$ as a population in the 2DEG layer, and
the contribution with $|g_2| \approx 0.43$ as a population
in a bulk GaAs layer.\cite{rizo2009prb} The slow growth
during delays till $\sim$100~ps (as compared to
mono-exponential decay for a single population) of signals
(at 0~T) and signal envelopes also originates from having a
superposition of two contributions. It occurs because the
amplitudes $A_i$ always have opposite sign (for $\phi_i$
around zero).\cite{rizo2009prb}

SDA is apparent in the zero-field Kerr signals in
Fig.~\ref{fig:sdazerofield}(a) from wires oriented along
the [110], \mbox{[-110]}, [100], and [010] (and for
reference from unpatterned area). The decay of signals
during delays $t>100$~ps clearly depends on the wire
orientation. The loss of Kerr signal is fastest for
\mbox{[-110]} oriented wires (only slightly slower than for
unpatterned area). Kerr signal decay is slowest for [110]
oriented wires, and results for wires in [100] and [010]
directions are in between these extreme cases and nearly
identical. The inset of Fig.~\ref{fig:sdazerofield}(a)
shows that the time scale for slow signal growth for
$t<100$~ps has a similar dependence on wire orientation,
with again the slowest time scale for the [110] direction.
For the Kerr signals measured in the range 1~T to 7~T in
Fig.~\ref{fig:sdainfield}(a), all traces have envelopes
with slow growth for $t<100$~ps, and decaying envelopes for
$t>100$~ps as the zero-field signals. At 1~T, the
corresponding time scales are again clearly longest for
[110] wires and shortest for \mbox{[-110]} wires but the
differences degrade for $B>1$~T. We only present signals
for the extreme cases \mbox{[-110]} and [110], results for
[100] and [010] wires show intermediate behavior.

Fitting these results with Eq.~(\ref{eq:kerrfitformula}) yields traces
which very closely match the experimental traces (therefore not
shown). Signals from $B \geq 1$~T give results for $\tau_i$ for both
the 2DEG ($|g_1|\approx0.36$) and bulk ($|g_2|\approx0.43$)
contributions to the signals [Figs.~\ref{fig:sdainfield}(b) and~\ref{fig:sdainfield}(c)]. This
identifies that the time scale for slow growth ($t<100$~ps) is
dominated by dephasing of the 2DEG population, while the later decay
of signals ($t>100$~ps) is dominated by dephasing of a bulk
population. The behavior for $B=0$~T
[Figs.~\ref{fig:sdazerofield}(b) and~\ref{fig:sdazerofield}(c)] is interpreted in the same way
from extrapolating down the trends in $\tau_i$ for $B>0$~T. This
identification of populations based on $g$ factors is consistent
with reported values for dephasing times in high-mobility 2DEGs
[$\sim$50~ps due to rapid D'yakonov-Perel' (DP) dephasing
\cite{miller2003prl,snelling1991prb,stich2007prl}] and intrinsic
GaAs layers. [$\sim$300~ps \cite{kikkawa1998prl}]


The results in Figs.~\ref{fig:sdazerofield}(b) and
\ref{fig:sdainfield}(b) give clear evidence for SDA for the 2DEG
population, with qualitative agreement with predicted behavior.\cite{liu2009jsnm,strain}
These predictions were based on the assumption
that Rashba and linear Dresselhaus SOI terms dominate. Expressed as
$k$-vector-dependent effective fields these are $\vec{B}_{R} =
C_{R}(\hat{x}k_{y}-\hat{y}k_{x})$ and $\vec{B}_{D1} =
C_{D1}(-\hat{x}k_{x}+\hat{y}k_{y})$, respectively, with typical
magnitudes of about 2~T for heterojunction 2DEGs
\cite{miller2003prl}. A summary of the numerical approach is given
below for the bulk population. The dependence of 2DEG dephasing on
wire orientation is here less strong than expected for exact
cancellation for $C_{D1}=C_{R}.$\cite{liu2009jsnm,koralek2009nature}
Instead, it is consistent with simulations for
$|C_{D1}+C_{R}|/|C_{D1}-C_{R}| \approx 3$ and $C_{D1}$ and $C_{R}$
as reported values.\cite{miller2003prl} We cannot make more
detailed statements about $C_{D1}$ and $C_{R}$ values since we
analyzed that this should account for the influence of the cubic
Dresselhaus SOI, momentum relaxation, and electron-hole recombination
on the observed Kerr signals. The dependence of SDA on $B$ is
discussed below.

The SDA observed for spins in a bulk layer in
Figs.~\ref{fig:sdazerofield}(c) and \ref{fig:sdainfield}(c)
was an unexpected result since SDA was until now only
considered for 2D electron systems. Apparently, our samples
also contain quasi-1D ensembles with bulk characteristics
that have anisotropy for SOI. Such ensembles must indeed be
present because band bending is significant up to
$\sim$2~$\mu$m below the wafer surface. Consequently, the
lifting of the conduction band below etched areas also
induces confining potentials along the wires at the depth
of the superlattice (which also drives a rapid drift of
photoelectrons from etched to unetched areas, such that
the bulk contribution to signals is mainly from unetched
regions). In addition, below the wires (areas not etched)
at the depth of the superlattice there is still a weak
slope in the conduction band [Fig.~\ref{fig:introfig}(c)].
A Rashba SOI should be effective here that only depends on
the $x$ and $y$ component of a $k$ vector as for the 2DEG
case. Here the Rashba parameter $C_{R} \propto \langle
dV/dz \rangle$ still has a substantial value \cite{rashba}
due to the slope of the conduction band ($V$ is potential
of the bottom of conduction band and $z$ is growth direction).
This results in anisotropic SOI [analogs to
Fig.~\ref{fig:introfig}(a)] when this Rashba term has a
similar magnitude as the Dresselhaus SOI for bulk.\cite{strain}

We therefore conjecture that the observed SDA for a bulk population
is due to electrons \textit{around} (\textit{i.e.}, not inside
\cite{rizo2009prb}) the superlattice. The Kerr reflection from the
superlattice will probe the spin imbalance in its vicinity. Most
likely, this population is located in the accumulation layer
[Fig.~\ref{fig:introfig}(c)] since the thickness of this layer is
sufficient for absorbing most of the energy from our laser pulses.
The separation into two distinct spin populations can then only be
explained if we assume the surprisingly low mobility for
photo-electrons along the growth direction that was recently
reported.\cite{salis2006prl} We can therefore not fully rule out
scenarios where the bulk response is dominated by electrons at the
substrate side of the superlattice, given that these low mobility
values are not yet fully understood.\cite{rizo2009prb}

To investigate the above scenario, we use a simple model
and Monte Carlo simulations as for 2DEG-based wires.\cite{liu2009jsnm}
We model a wire of bulk material by
assuming a randomly moving electron ensemble with a three-dimensional
$k$ vector distribution that is confined to a rectangular
elongated box with short sides of 1~$\mu$m. The electrons
are scattering on the edge of the box and on impurities
(mean-free path from impurities alone set at 1~$\mu$m,
\textit{i.e.}, a quasiballistic electron ensemble). The
(quasi-)Fermi level is derived from the estimated
photo-electron density around the superlattice ($\sim 3
\cdot 10^{21} \; \rm{m}^{-3}$). Spins are precessing around
the sum of anisotropic spin-orbit fields and the external
field. We assume the electron spins experience only the
bulk Dresselhaus SOI Ref.(\onlinecite{silsbee2004jpcm}), and the
mentioned Rashba SOI. Dephasing is again purely due to the
DP mechanism. Such numerics indeed produces dephasing times
similar to the observed values. Also, it shows SDA as a
function of wire orientation with the slowest dephasing for
wires along [110] and with the ratio between $\tau_2$ for
[110] and \mbox{[-110]} wires close to 3 for $B=0$~T.

As a final point we discuss the magnetic field dependence
of SDA for the 2DEG and bulk populations. For both, the SDA
(taken as the ratio between dephasing times for [110] and
\mbox{[-110]} wires, see Figs.~\ref{fig:sdainfield}(b) and ~\ref{fig:sdainfield}(c))
is most apparent at zero external field, and degrades with
increasing field. Our type of SDA only occurs in quasi-1D
systems, \textit{i.e.}, when the width of the wires is less
than the spin precession length $L_{sp}$.\cite{liu2009jsnm}
Alternatively, it can degrade when an
external field starts to dominate over anisotropic
spin-orbit fields [Fig.~\ref{fig:introfig}(a)]. The
degradation of SDA of the bulk population is consistent
with both explanations. The external field drives $L_{sp}$
below $\sim$1~$\rm{\mu m}$ for $B\approx 4$~T (using again
the estimated density of photo-electrons around the
superlattice) but we also estimate \cite{miller2003prl}
that the external field dominates over spin-orbit fields
above $B \approx 0.3$~T. For the 2DEG a much stronger field
is needed for driving $L_{sp} \lesssim 1 \; \rm{\mu m}$,
but the external field dominates over spin-orbit fields for
$B> 4$~T \cite{miller2003prl}. We repeated all measurements
with the external field in \mbox{[-110]} direction. We
observed small shifts ($\sim10\%$) in the dephasing times,
but the $B$ dependence of SDA was at nearly the same level.
We interpret this as follows. Having only partial
cancellation of SOI (not the pure case $C_{D1}=C_{R}$) gives
anisotropy for SOI with fields varying both in magnitude
and direction. This still gives SDA, but it can be lifted
by adding an external field that is stronger than the
spin-orbit fields, and there is then little dependence on
the direction of the field. Our Monte-Carlo simulations
confirm this behavior, and give SDA values that degrade
from $\sim$3 at $B=0$~T to $\sim$2 at 7~T.

In conclusion, our results confirm that confining electron
ensembles in quasi-1D wires in [110] direction can yield
suppressed spin dephasing. The results agree with
predictions based on anisotropy for SOI. Further studies
with tuning of SOI by a gate gives access to better
cancellation of Rashba and Dresselhaus SOI, and can explore
the ultimate limit of suppressing dephasing in wires
\cite{koralek2009nature,liu2009jsnm}. Such gate control can
also yield spin-transistor functionality
\cite{awschalom2009phys} by associating \textit{on} and
\textit{off} with strong or little dephasing of spin
signals. Our results also establish that using SDA for
suppressing dephasing is not restricted to quantum-well
electron ensembles. It is also applicable to quasi-1D
ensembles in bulk-like layers where the conduction band has
a slope. Our results thus open the path to investigating
the mentioned spin-transistor functionality for bulk
semiconductor channels \cite{dash2009nature}.

We thank R.~Winkler and A.~Rudavskyi for discussions, and
the Dutch NWO and NanoNed, and the German programs DFG-SFB
491, DFG-SPP 1285 and BMBF nanoQUIT for financial support.


\begin{thebibliography}{99}
\bibitem{grating} We checked that grating behavior of
the arrays did not influence Kerr signals, by rotating the
polarization planes of the pump and probe beam around their
axes.
\bibitem{strain} We analyzed that SOI due to possible
presence of wire-induced strain cannot explain the observed SDA.
\bibitem{rashba} As compared to the 2DEG case, where the slope in V for the
GaAs layer is compensated by the steep increase in V at the
heterojunction, since 2DEG states penetrate the AlGaAs
layer (Fig.~\ref{fig:introfig}(c)).
\end{thebibliography}
\end{document}